\documentclass[aps,prd,preprintnumbers,groupedaddress,nofootinbib]{revtex4}

\pdfoutput=1

\usepackage{graphicx}
\usepackage{amsfonts}
\usepackage{amssymb}
\usepackage{amsmath}
\usepackage{slashed}
\usepackage{url}
\usepackage{multirow}
\usepackage{color}

\begin{document}
\def\tablename{Table}
\def\figurename{Figure}

\def\gtot{\Gamma_\text{tot}}
\def\brinv{\text{BR}_\text{inv}}
\def\brsm{\text{BR}_\text{SM}}
\def\bratio{\mathcal{B}_\text{inv}}
\def\as{\alpha_s}
\def\az{\alpha_0}
\def\gz{g_0}
\def\w{\vec{w}}
\def\sdag{\Sigma^{\dag}}
\def\s{\Sigma}
\newcommand{\psib}{\overline{\psi}}
\newcommand{\Psib}{\overline{\Psi}}
\newcommand\one{\leavevmode\hbox{\small1\normalsize\kern-.33em1}}
\newcommand{\Mpl}{M_\mathrm{Pl}}
\newcommand{\p}{\partial}
\newcommand{\lag}{\mathcal{L}}
\newcommand{\qqquad}{\qquad \qquad}
\newcommand{\qqqquad}{\qquad \qquad \qquad}

\newcommand{\qb}{\bar{q}}
\newcommand{\matx}{|\mathcal{M}|^2}
\newcommand{\really}{\stackrel{!}{=}}
\newcommand{\msbar}{\overline{\text{MS}}}
\newcommand{\qns}{f_q^\text{NS}}
\newcommand{\lqcd}{\Lambda_\text{QCD}}
\newcommand{\met}{\slashchar{p}_T}
\newcommand{\pmiss}{\slashchar{\vec{p}}_T}

\newcommand{\st}[1]{\tilde{t}_{#1}}
\newcommand{\stb}[1]{\tilde{t}_{#1}^*}
\newcommand{\nz}[1]{\tilde{\chi}_{#1}^0}
\newcommand{\cp}[1]{\tilde{\chi}_{#1}^+}
\newcommand{\cm}[1]{\tilde{\chi}_{#1}^-}

% all the masses 
\providecommand{\mg}{m_{\tilde{g}}}
\providecommand{\mst}{m_{\tilde{t}}}
\newcommand{\msn}[1]{m_{\tilde{\nu}_{#1}}}
\newcommand{\mne}[1]{m_{\tilde{\chi}_{#1}}}
\newcommand{\msb}[1]{m_{\tilde{b}_{#1}}}

% units of measure
\newcommand{\mev}{{\ensuremath\rm MeV}}
\newcommand{\gev}{{\ensuremath\rm GeV}}
\newcommand{\tev}{{\ensuremath\rm TeV}}
\newcommand{\fb}{{\ensuremath\rm fb}}
\newcommand{\ab}{{\ensuremath\rm ab}}
\newcommand{\pb}{{\ensuremath\rm pb}}
\newcommand{\sign}{{\ensuremath\rm sign}}
\newcommand{\ifb}{{\ensuremath\rm fb^{-1}}}

% really great macro by Chris Lester
\def\slashchar#1{\setbox0=\hbox{$#1$}           % set a box for #1
   \dimen0=\wd0                                 % and get its size
   \setbox1=\hbox{/} \dimen1=\wd1               % get size of /
   \ifdim\dimen0>\dimen1                        % #1 is bigger
      \rlap{\hbox to \dimen0{\hfil/\hfil}}      % so center / in box
      #1                                        % and print #1
   \else                                        % / is bigger
      \rlap{\hbox to \dimen1{\hfil$#1$\hfil}}   % so center #1
      /                                         % and print /
   \fi}
\newcommand{\dslash}{\slashchar{\partial}}
\newcommand{\Dslash}{\slashchar{D}}

\def\eg{{\sl e.g.} \,}
\def\ie{{\sl i.e.} \,}
\def\etal{{\sl et al} \,}

\title{Stop on Top}

\author{Matthew R.~Buckley$^{1}$, Tilman Plehn$^{2}$, and Michael J.~Ramsey-Musolf$^{3,4}$}
\affiliation{$^1$ Department of Physics and Astronomy, Rutgers University, Piscataway, USA}
\affiliation{$^2$ Institut f\"ur Theoretische Physik, Universit\"at Heidelberg, Germany}
\affiliation{$^3$ Amherst Center for Fundamental Interactions, Department of Physics, University of Massachusetts, Amherst, USA}
\affiliation{$^4$ Kellogg Radiation Laboratory, California Institute of Technology, Pasadena, CA USA}

\preprint{ACFI-T14-05}
%\date{\today}

\begin{abstract}
The most natural supersymmetric solution to the hierarchy problem
prefers the scalar top partner to be close in mass to the top
quark. Experimental searches exclude top squarks across a wide range
of masses, but a gap remains when the difference between the masses of the stop 
and the lightest supersymmetric particle is close to the top mass.  We
propose to search for stops in this regime by exploiting the azimuthal angular correlation
of forward tagging jets in (s)top pair production. As shown in earlier work, this correlation
is sensitive to the spin of the heavy states, allowing one to distinguish between top and stop pair production.
Here, we demonstrate that this angular
information can give a statistically significant stop pair production signal in the upcoming LHC run. 
While the appropriate simulation including parton showering and detector simulation
requires some care, we find stable predictions for the angular
correlation using multi-jet merging.
\end{abstract}

\maketitle

\bigskip \bigskip \bigskip
\tableofcontents

\newpage

%%%%%%%%%%%%%%%%%%%%%%%%%%%%%%%%%%%%%%%%%%%%%%%%%%%%%%%%%%%%%%%%%%%%%%%%%%%%%%%
\section{Introduction}
\label{sec:intro}

The top partner holds a special place in many extensions of the
Standard Model~\cite{bsm_review}.  As the fermion with the
largest coupling to the Higgs field, the top gives the largest
quadratic correction to the Higgs mass term. To have a natural and
untuned cancellation of this term, we would expect the supersymmetric
top squark --- the stop (${\tilde t}$) --- to be close in mass to the top itself.
Additionally, in generic supersymmetric flavor models the large top Yukawa
drives the mixing of left-- and right--handed stops and pushes the
lightest stop mass eigenstate to be the lightest squark. Experimentally, however,
no evidence of a relatively light stop has been obtained in collider searches.
A combination of 
ATLAS~\cite{atlas_stops}
and
CMS~\cite{cms_stops}
results at 7 and 8 TeV excludes stop pair production decaying to final
states containing an invisible, stable supersymmetric particle ({\em e.g.}, the lightest neutralino, $\tilde{\chi}^0$) for stop
masses in the range of $100- 750$~GeV, assuming a massless invisible
decay product.\bigskip

Nevertheless, in the two-dimensional plane of ${\tilde t}$ and $\nz{}$ masses, there remains a
notable window in the experimental exclusion regions: neither experiment has ruled out the
possibility that stop pair production events may be buried top in production  when the mass difference
$\mst -(\mne{} + m_t)$ becomes small.
There is a simple explanation for this lack of sensitivity to stop 
production near the ``degeneracy line:'' when the mass
splitting is small, the invisible particles ($\nz{}$) carry  little momentum,
so the final state from stop pair production closely mimics 
that of top pairs in the Standard Model. In principle, measurable differences in the
missing transverse energy ($\slashed{E}_T$) distributions would for fully hadronic top decays would appear if
stop events are also present, a feature that might allow discovery or exclusion of degenerate stops~\cite{degenerate_stops_had}.
In practice, however, such searches face challenging jet combinatorics and require
precise understanding of the background $\slashed{E}_T$. In the di- or semi-leptonic channels, kinematic variables built from the decay
products of the top are nearly identical for $t\bar{t}$
and $\st{} \st{}^*$ events, assuming the stop decays to either (a) an on-shell
or off-shell top and an invisible $\nz{}$, or (b) a bottom quark and a
chargino, where the latter decays into a $\nz{}$ and a $W^{(\ast)}$ boson.  Analyzing differences in the top production angles or top
decay products have been suggested~\cite{more_light_stops} to
search for stop pairs contaminating the top sample, but the possible
improvement is small and can be washed out by necessary trigger and
selection criteria.\bigskip

In this study we
explore an alternative approach for distinguishing top and stop pair production that avoids these difficulties. Specifically, 
we show how correlations between tagging jets can be used to search
for stop pairs in the top pair sample at the
LHC~\cite{kaoru} independent of the stop decays. In particular, we consider the difference in the
azimuthal angles $\Delta\phi$ of forward jets produced in association
with the top or stop pair in vector boson fusion (VBF)
events.\footnote{Here, the fusing vector bosons are primarily gluons,
  justifying the term ``VBF.''}  These jets arise from initial state
radiation. The information in their $\Delta \phi$ distribution can be
used regardless of decay channels, as long as we can manage to extract a
signal-rich sample. As was originally demonstrated in the context of
Higgs
physics~\cite{delta_phi,higgs_spin},
the difference in azimuthal angle between the two forward jets $\Delta
\phi$ from weak--boson--fusion events inherit information about the
helicities of the weak bosons involved in the production. From the
underlying argument it is obvious that this technique can be generalized to gluon
fusion~\cite{higgs_spin,delta_phi_gg}.  The helicities that can participate in a
given process are set by the Lorentz structure of the production
matrix element, and so for pair production the distribution of
$\Delta\phi$ is sensitive to properties of pair-produced particles such as
spin and CP assignment.\bigskip

For the pair production process of interest here, the resulting differential cross section has the form
\begin{equation}
\label{eq:delphidist}
\frac{d\sigma}{d\Delta\phi} = 
A_0 + A_1 \cos \Delta\phi+A_2 \cos (2\Delta\phi)\ \ \ ,
\end{equation}
where the expansion coefficients $A_k$ encode the interplay of the underlying pair production
amplitude and the helicity of the fusing gluons. As shown in our earlier work~\cite{matt_michael},  
the sign of $A_2$ is set by the spins of the produced particles: $A_2>0$ for scalars and $A_2<0$ 
for fermions. In general, this sensitivity could provide a powerful technique for
diagnosing the spin of  any new particles that may be discovered at the LHC~\cite{matt_michael}. 
This is also the case for top pair production close to threshold, while in the
relativistic limit the sum of the two azimuthal angles is the more
sensitive observable~\cite{kaoru}. 
In the present context, we show how one may exploit the same effect to identify or exclude the presence
of stop pairs in the region of  parameter space near the degeneracy line. Moreover,
we describe how the $\cos (2\Delta\phi)$ correlation between initial state radiation jets
can be reliably described in event simulations that take into account parton showering and realistic
detector jet identification and show that the correlation is not washed 
out through azimuthal decorrelation \cite{daCosta:2011ni,Khachatryan:2011zj}  To our knowledge, this study represents
the first such demonstration, indicating that study of azimuthal tagging jet correlations may be a realistic
tool in other contexts as well~\cite{matt_michael,kaoru}. \bigskip

Before determining if the degenerate stop production could be hiding in top pair production at the LHC, one should ask whether the measured cross section for top
pair production allows for such a scenario.  This rate has been measured
numerous
times~\cite{atlas_top_0,cms_top_0,atlas_top_1,cms_top_1,tevatron_top}
and agrees with theoretical predictions~\cite{Czakon:2013goa} within
uncertainties. In Table~\ref{tab:xsection}, we show the measured top
pair cross sections at the Tevatron and the LHC, along with the
theoretical predictions and the supersymmetric stop pair production
cross sections for light stop masses of 175 and 200~GeV. At first glance,
the measured cross section would appear to rule out the addition
of a stop with mass near that of the top. However, it is unclear how
the top cross section measurements would respond to an admixture of
stop events, and there may be a degeneracy between the cross section
and top mass measurements. Short of a detailed analysis of this question that goes beyond the scope of the present study, we cannot rule out the possibility -- however unlikely -- that a 175~GeV stop could be hiding inside the top sample. Moreover, a stop with mass around 200~GeV, still within the degeneracy window, is not in significant tension with the
experimental results, given the uncertainties. Consequently, we will consider two benchmark cases, corresponding  to 
$(\mst,\mne{})=(175,1)$ GeV and $(200, 25)$ GeV, respectively.

%---------------------------------------------
\begin{table}[t]
\begin{tabular}{c|c|c|c|c}
\hline
$\sqrt{s}$~[TeV] & $\sigma_{t\bar{t}}~$[pb] & $\sigma_{t\bar{t}}~$[pb] & $\sigma_{\st{} \st{}^*}$~[pb] & $\sigma_{\st{} \st{}^*}$~[pb] \\ & experiment & theory & $\mst = 175~\gev$ &  $\mst = 200~\gev$ \\ \hline
1.96 & $7.68\pm0.20_\text{stat}\pm0.36_\text{sys}$~(CDF+D\O\ \cite{tevatron_top}) 
     & $7.164{^{+0.110}_{-0.200}}_\text{scale}{^{+0.169}_{-0.122}}_\text{pdf}$ & 0.587 & 0.252 \\ \hline
7 & $\begin{array}{c} 177\pm3_\text{stat} \pm {^8_7}_\text{sys}\pm7_\text{lumi}~\text{(ATLAS)} \\ \pm3_\text{stat} \pm {^8_77}_\text{sys}\pm7_\text{lumi}~\text{(CMS)} \end{array}$
  & $172.0{^{+4.4}_{-5.8}}_\text{scale}{^{+4.7}_{-4.8}}_\text{pdf}$ & 24.0 & 11.9 \\ \hline
8 & $\begin{array}{c} 238\pm2_\text{stat} \pm 7_\text{sys}\pm7_\text{lumi}\pm 4_{\text{beam~$E$}}~\text{(ATLAS~\cite{atlas_top_1})} \\ 227 \pm 3_\text{stat} \pm 11_\text{sys}\pm 10_\text{lumi}~\text{(CMS \cite{cms_top_1})} \end{array}$
  & $245.8{^{+6.2}_{-8.4}}_\text{scale}{^{+6.2}_{-6.4}}_\text{pdf}$ & 34.5 & 17.3 \\ \hline
14 & -- & $953.6{^{+22.7}_{-33.9}}_\text{scale}{^{+16.2}_{-17.8}}_\text{pdf}$ & 135 & 72.1 \\ \hline
\end{tabular}
\caption{Cross sections for top and stop pair production at the 1.96
  TeV Tevatron and 7, 8, and 14 TeV LHC. The theoretical predictions
  for the $t\bar{t}$ cross sections are calculated at NNLO+NNLL, for
  $m_t = 173.3$~GeV~\cite{Czakon:2013goa}. Cross sections for stop
  pair production are calculated at NLO in {\tt
    Prospino2}~\cite{prospino}
  with a light $\st{1}$ and all other supersymmetric particles
  decoupled.}
\label{tab:xsection}
\end{table}
%---------------------------------------------

Our discussion is organized as follows. In Section~\ref{sec:spin} we explain the physics behind the
$\Delta\phi$ correlations of VBF tagging jets in the specific cases of
top and stop pair production. In Section~\ref{sec:simulation} we then
discuss the simulation of these events including multi-jet merging in
{\tt MadGraph5}. While in the default setup the correlations between
the tagging jets are not guaranteed to be included we show how they can be accounted
for.  In the same section we study the tagging jet correlations at
parton level and show how a dedicated analysis can separate top and
stop contributions to a mixed event sample.  In
Section~\ref{sec:searches} we confirm that using realistic cuts and a
fast detector simulation these results can be reproduced.

%%%%%%%%%%%%%%%%%%%%%%%%%%%%%%%%%%%%%%%%%%%%%%%%%%%%%%%%%%%%%%%%%%%%%%%%%%%%%%%
\section{Tagging jet correlations}
\label{sec:spin}

\begin{figure}[b!]
\includegraphics[width=0.23\textwidth]{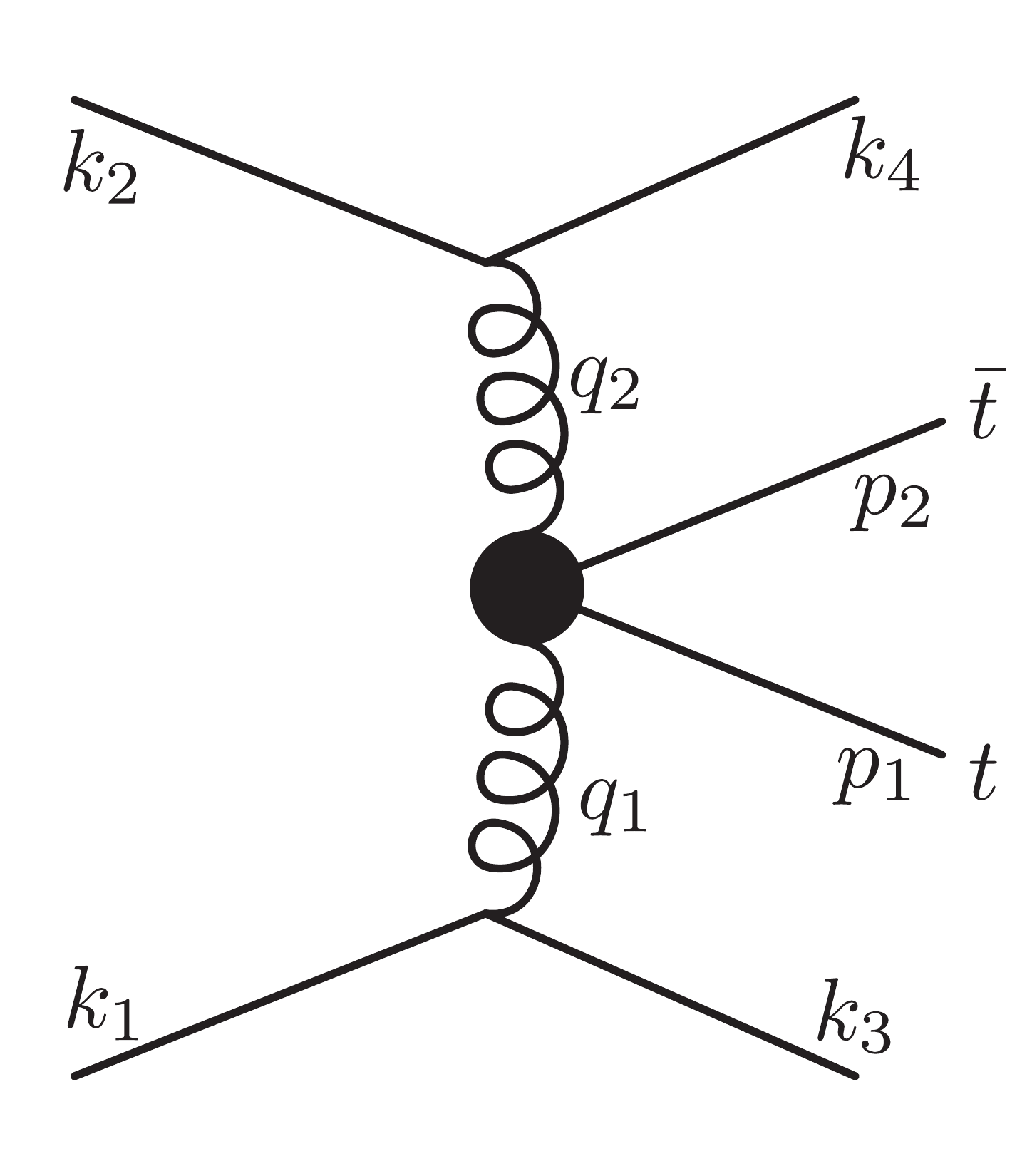}
\caption{A representative Feynman diagram for the VBF process $pp\to
  t\bar{t}+j j$ with two tagging jets. Similar diagrams exist for stop
  pair production. The initial and final state partons can be quarks,
  anti-quarks, or gluons. The different channels contributing to the
  hard $gg \to t\bar{t}$ scattering are denoted by a solid dot.}
\label{fig:feynman}
\end{figure}

\begin{figure}[t]
 \includegraphics[width=0.6\textwidth]{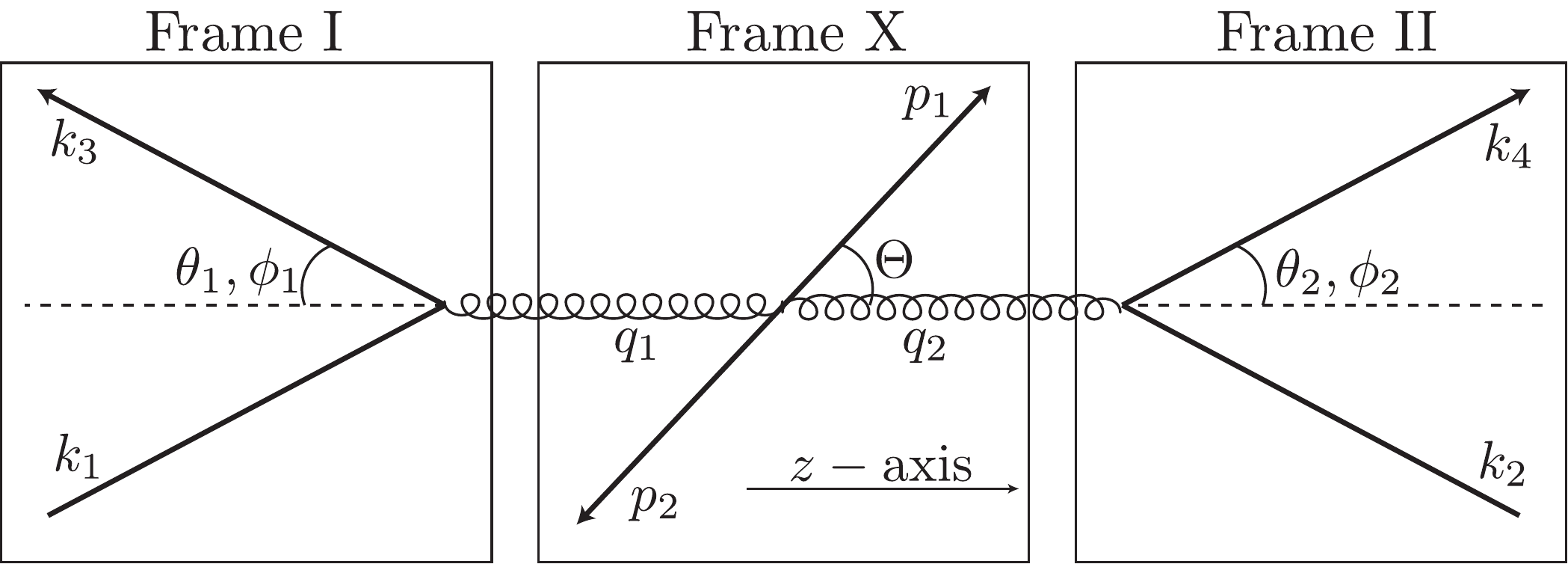}
\caption{Kinematics for VBF events, showing the two Breit frames~I and
  II and the production frame~X~\cite{delta_phi}.}
\label{fig:kinematics}
\end{figure}

We are interested in top and stop pairs with two associated tagging
jets, produced primarily via initial state radiation, or equivalently,
through VBF
diagrams~\cite{tagging}. Eventually, to separate
VBF production from all other sources of jets we will
employ strict selection cuts, primarily requiring the jets to be
forward. A representative Feynman diagram is shown in
Figure~\ref{fig:feynman}, defining our notation for the different
momenta. The full gauge-invariant matrix element will be the sum of
many diagrams, but the cuts will emphasize this topology's contribution
to the amplitude. In our simulations, we will
include all initial parton states, though in practice gluons dominate for
the parameter range of interest.
It is most convenient to write the relevant kinematics
in the three frames shown in
Figure~\ref{fig:kinematics}~\cite{delta_phi}. The emission of
the fusing vector bosons (gluons in our case) from the incoming
partons are described in the Breit frames (frames~I and II), defined
by the gluon momenta being purely space--like and in the
$z$-direction:
\begin{alignat}{5}
q_1^\mu & = k_1^\mu - k_3^\mu = (0,0,0,Q_1), \notag \\
q_2^\mu & = k_2^\mu - k_4^\mu = (0,0,0,-Q_2) \; .
\end{alignat}
The top/stop pair production frame shown as frame~X in
Figure~\ref{fig:kinematics} is defined as the frame in which
$q_1^\mu+q_2^\mu = (\sqrt{\hat{s}},\vec{0})$, where $\hat{s} \equiv
(p_1+p_2)^2$ is the invariant mass of the top or stop
pair.\bigskip

%---------------------------------------------
%---------------------------------------------

%---------------------------------------------
%---------------------------------------------

We now focus on the dependence of the differential cross section on
the azimuthal angles $\phi_1$ and $\phi_2$.  As long as the tagging
jets with the momenta $k_3$ and $k_4$ are forward, the $z$-axis shared by frames~I, II,
and X is nearly collinear with the experimental beam axis. As a first
step we can approximate the observed azimuthal angles in the
laboratory frame by the angles in the plane orthogonal to the top or
stop momenta~\cite{higgs_spin}. The matrix element for the full
VBF event takes the form
\begin{equation}
{\cal M} = \sum_{h_1,h_2} 
{\cal M}_\text{I}^\mu(h_1,\phi_1,\theta_1)
{\cal M}_\text{II}^\nu(h_2,\phi_2,\theta_2)
{\cal M}_\text{X}^{\mu\nu}(h_1,h_2,\Theta) \; ,
\end{equation}
where $h_1,h_2=-1,0,+1$ are the helicities of the gluons $q_1$ and
$q_2$, measured relative to the $z$-axis, so $h_1 = +1$ is positive
angular momentum for $q_1$, but $h_2 = -1$ is positive angular
momentum for $q_2$. We suppress the dependence on the color factors. A
boost is required to take each matrix element from its individual
frame to a common center--of--mass frame. All these boosts will be in
$z$-direction and will not induce additional dependence on the
azimuthal angles $\phi_i$. Therefore, $\phi_1$ and $\phi_2$ enter
only as phases of the Breit matrix elements,
\begin{alignat}{5}
{\cal M}_\text{I}(h_1,\phi_1,\theta_1) &= 
{\cal M}_\text{I}(h_1,0,\theta_1) \; e^{+ih_1\phi_1}, \notag \\
{\cal M}_\text{II}(h_2,\phi_2,\theta_2) &= 
{\cal M}_\text{II}(h_2,0,\theta_2) \; e^{-ih_2\phi_2}.
\end{alignat}
We can rewrite $\phi_1$ and $\phi_2$ in terms of their difference
$\Delta \phi \equiv \phi_1-\phi_2$ and their sum $\phi_+ \equiv
\phi_1+\phi_2$~\cite{delta_phi,higgs_spin}. The angle $\phi_+$ is
physically unobservable without reference to the top or stop
production plane, which we will not attempt to reconstruct, and so it
can be integrated over. Abbreviating the six-body phase space factors as
$({\cal PS})$ and the integration over all other angles as $d\Omega$,
the differential cross section with respect to $\Delta\phi$ can be
written as
\begin{equation}
\frac{d\sigma}{d\Delta \phi} = ({\cal PS})  \int d\Omega\sum_{h_1^{(')},h_2^{(')}} 
e^{i\Delta h \; \Delta\phi/2}
\left[{\cal M}_\text{I}^\mu(h_1){\cal M}_\text{I}^{\mu'*}(h_1')\right]
\left[{\cal M}_\text{II}^\nu(h_2){\cal M}_\text{II}^{\nu'*}(h_2')\right]
\left[{\cal M}_\text{X}^{\mu\nu}(h_1,h_2){\cal M}_\text{X}^{\mu'\nu'*}(h_1',h_2') \right] \; ,
\end{equation}
with $\Delta h = h_1-h_1'+h_2-h_2'$.  This distribution has to be
invariant under the shift $\Delta \phi \to \Delta \phi+2\pi$, which
translates into the condition $\Delta h = 0, \pm 2, \pm 4$.  Terms
with odd $\Delta h$ must vanish, and larger values of $\Delta h$
cannot be generated for $|h_j| \le 1$ (allowing for
off-shell gluons).  We then expand the exponential
with the helicities in sines and cosines and, assuming CP conservation,
ignore the complex sine contributions. The three allowed helicity
changes $\Delta h$ give rise to the three coefficients of
Eq.~(\ref{eq:delphidist}),
\begin{alignat}{5}
A_n & =  ({\cal PS}) \int d\Omega \sum_{\Delta h = \pm n} 
\left[{\cal M}_\text{I}^\mu(h_1){\cal M}_\text{I}^{\mu'*}(h_1')\right]
\left[{\cal M}_\text{II}^\nu(h_2){\cal M}_\text{II}^{\nu'*}(h_2')\right]
\left[{\cal M}_\text{X}^{\mu\nu}(h_1,h_2){\cal M}_\text{X}^{\mu'\nu'*}(h_1',h_2') \right] \; .
\label{eq:diffsigma}
\end{alignat}
%
%with 
%
%\notag \\
%A_1 & =  ({\cal PS})  \int d\Omega\sum_{\Delta h = \pm 2} 
%\left[{\cal M}_\text{I}^\mu(h_1){\cal M}_\text{I}^{\mu'*}(h_1')\right]
%\left[{\cal M}_\text{II}^\nu(h_2){\cal M}_\text{II}^{\nu'*}(h_2')\right]
%\left[{\cal M}_\text{X}^{\mu\nu}(h_1,h_2){\cal M}_\text{X}^{\mu'\nu'*}(h_1',h_2') \right] \notag \\
%A_2 & =  ({\cal PS})  \int d\Omega\sum_{\Delta h = \pm 4} 
%\left[{\cal M}_\text{I}^\mu(h_1){\cal M}_\text{I}^{\mu'*}(h_1')\right]
%\left[{\cal M}_\text{II}^\nu(h_2){\cal M}_\text{II}^{\nu'*}(h_2')\right]
%\left[{\cal M}_\text{X}^{\mu\nu}(h_1,h_2){\cal M}_\text{X}^{\mu'\nu'*}(h_1',h_2') \right] \; .
%\end{alignat}
%
We will be most interested in $A_2$, where $\Delta h= \pm 4$. This can only
be satisfied by the unique configuration $h_1 = h_2 = \pm 1$ and $h_i'
= -h_i$.\bigskip

From explicit calculation, the contribution from the matrix
elements for gluon emission, \ie${\cal M}_\text{I}(h_1)^\mu{\cal
  M}_\text{I}(-h_1)^{\mu'*}$ and ${\cal M}_\text{II}(h_2)^\nu{\cal
  M}_\text{II}(-h_2)^{\nu'*}$ for $h_i = \pm 1$, are all
positive~\cite{delta_phi}. As a result, the sign of $A_2$
depends only on the sign of the pair production interference terms
${\cal M}_\text{X}^{\mu\nu}(h,h){\cal M}_\text{X}^{\mu'\nu'*}(-h,-h)$,
with $h = \pm 1$. That is, the sign of $A_2$ depends on the relative
sign between the matrix element for pair production where the total
incoming $z$-component of angular momentum is $+2$, and the matrix
element where the incoming $J_z = -2$ .

An explicit calculation of these interference terms in the case of the
fusion of abelian gauge bosons shows that, for the production of
scalars, these interference terms are overall positive, while for
fermion production, the terms are overall
negative~\cite{matt_michael}. We can now repeat this calculation in
the case of QCD-coupled heavy quarks~\cite{kaoru} or squarks. The
results are made more clear by multiplying the matrix elements in
frame~X by polarization vectors for the virtual gluons $q_1$ and
$q_2$, treating them as approximately on-shell. Recalling that
positive helicity for both gluons is defined relative to the $z$-axis,
rather than relative to the gluon momentum, both sets of polarization
vectors can be written as $\epsilon_{1/2}^\pm = 
(0,1,\pm i,0)/\sqrt{2}$.\bigskip

We begin with the fermionic case. For top pairs, the relevant
production matrix elements times polarization vectors in frame~X are
\begin{alignat}{5}
\left[{\cal M}^{\mu\nu} _\text{X}(h,h)\right]^{s,s} \epsilon_\mu(h)\epsilon_\nu(h) = & 
- \; \; \;  ig_s^2 \; 2s \;   
\left( \{T^a,T^b\}+\beta\cos\Theta [T^a,T^b] \right) \; 
\beta \sqrt{1-\beta^2} \; \frac{\sin^2\Theta}{1-\beta^2\cos^2\Theta} \notag \\
\left[{\cal M}^{\mu\nu} _\text{X}(h,h)\right]^{s,-s} \epsilon_\mu(h)\epsilon_\nu(h) = & 
- h \; ig_s^2 \; 2s \; 
\left(\{T^a,T^b\}+\beta\cos\Theta [T^a,T^b] \right) \; 
\beta \qquad \sin\Theta \frac{1- 2s h\cos\Theta}{1-\beta^2\cos^2\Theta} \; . 
\label{eq:fermionspin}
\end{alignat}
The angle $\Theta$ is defined in Figure~\ref{fig:kinematics}.  The
superscripts $s,s$ or $s,-s$ for $s= \pm 1/2$ denote the helicities of
the top and anti-top, measured relative to each of their momenta. In terms of the total
production energy $\hat{s}$ the
velocity of the top and anti-top $\beta$ is $\beta= \sqrt{1-4m^2/\hat{s}}$.

Notably, the matrix elements for production of a $t\bar{t}$ pair with
the same helicity assignments Eq.~\eqref{eq:fermionspin} do not have
the property that ${\cal M}_X(+1,+1) \times {\cal M}_\text{X}(-1,-1)^*
< 0$, contrary to our expectations. However, the signs of the $s,-s$
matrix elements with opposite helicity are manifestly asymmetric, as ${\cal
  M}_\text{X}^{s,-s}(h,h) \propto h$, so this product is indeed
negative. The fact that one term is not clearly negative could be
concerning for our argument, but by inspection it is clear that the negative
terms are strictly larger in magnitude than the positive
contributions. It is possible that the $\beta$ dependence of the
$A_2$ term could be useful in an experimental analysis. Cuts placed on
the top decay products could be used to enhance particular ranges of $\beta$~\cite{kaoru},
enhancing or suppressing the interference effect and providing useful 
side-bands. We will not further investigate this possibility in this paper. 
\bigskip

Turning to the stop pair production, the relevant matrix elements are
\begin{equation}
{\cal M}_\text{X}^{\mu\nu}(h,h)\epsilon_\mu(h)\epsilon_\nu(h) = 
ig_s^2 \; 
\left(\{T^a,T^b\}+\beta\cos\Theta [T^a,T^b] \right) \; 
\frac{\beta^2\sin^2\Theta}{1-\beta^2\cos^2\Theta} \; .
\label{eq:scalarM}
\end{equation}
Clearly this does not depend on the gluon helicities $h$, and so the
interference terms are positive. This results in a positive $A_2$ term
for stop pair production, and thus, the sign of $A_2$ can be used to
distinguish the production of scalar stops and fermionic tops. Note
that these two calculations only demonstrate that the top and stop
distributions will have opposite signs of their $A_2$ components,
without addressing the relative magnitudes. To answer that question,
we must turn to Monte Carlo simulation.

%%%%%%%%%%%%%%%%%%%%%%%%%%%%%%%%%%%%%%%%%%%%%%%%%%%%%%%%%%%%%%%%%%%%%%%%%%%%%%%
\section{Simulating VBF (S)Tops}
\label{sec:simulation}

%---------------------------------------------
\begin{figure}[b!]
\includegraphics[width=0.245\textwidth]{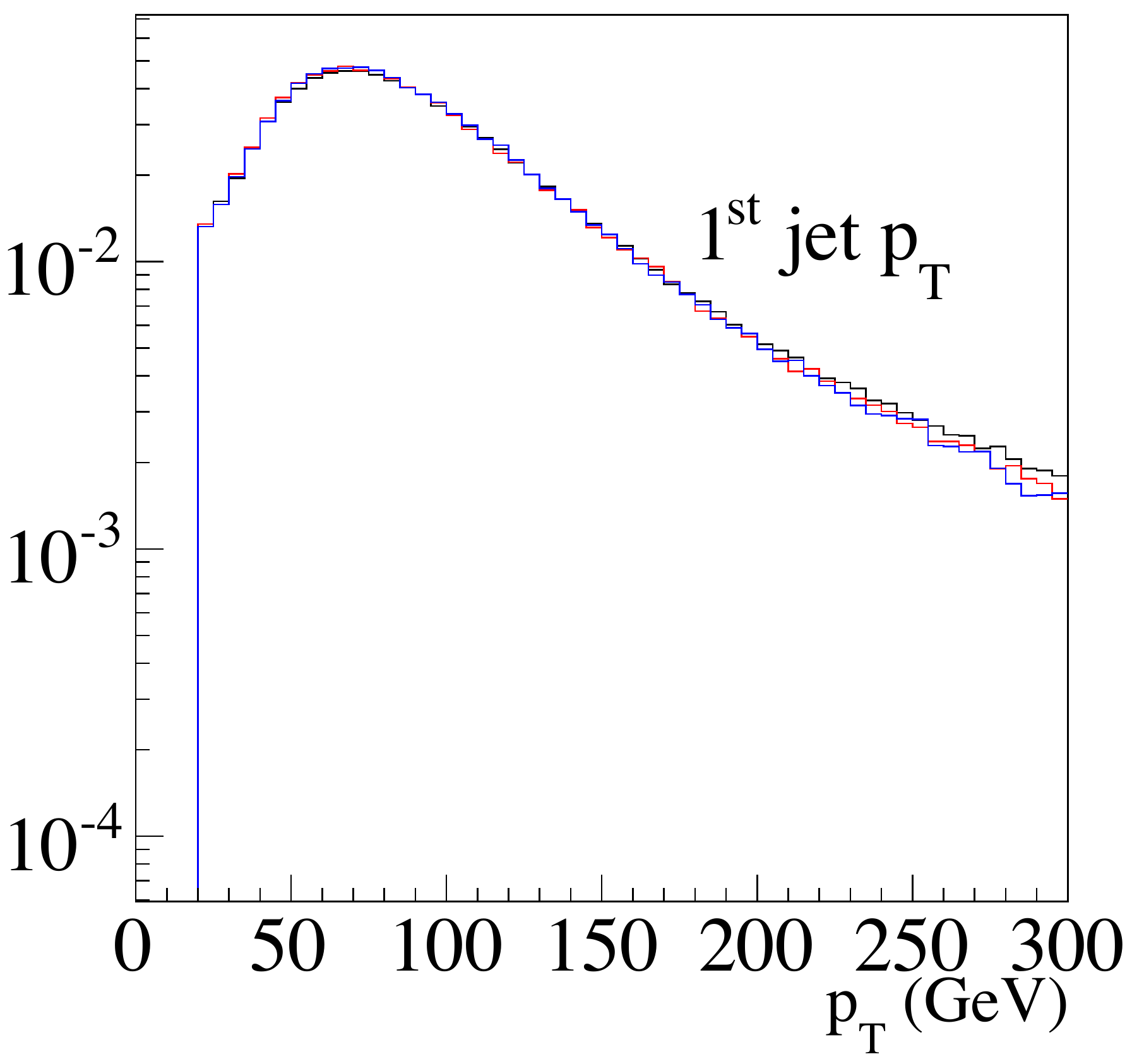}
\includegraphics[width=0.245\textwidth]{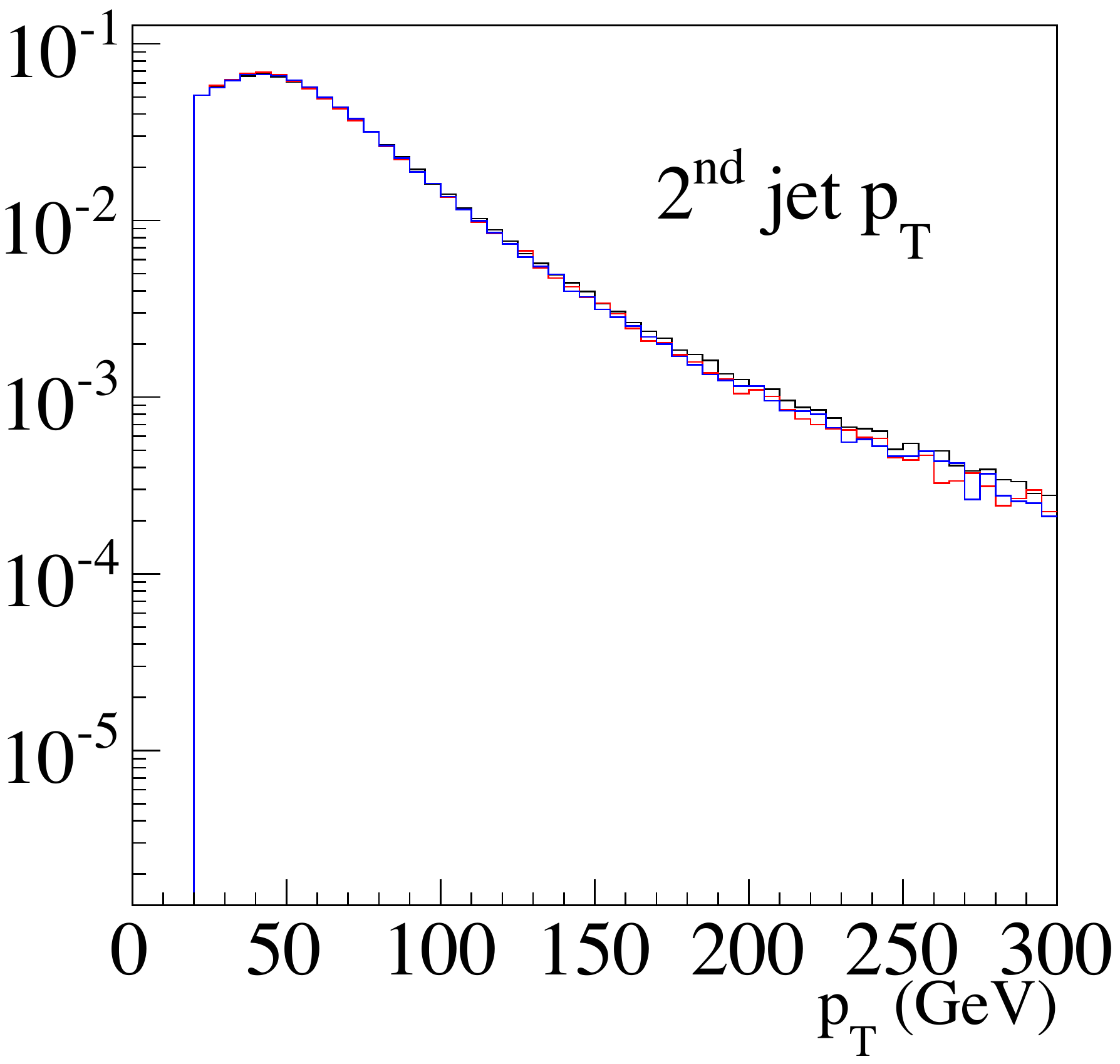}
\includegraphics[width=0.245\textwidth]{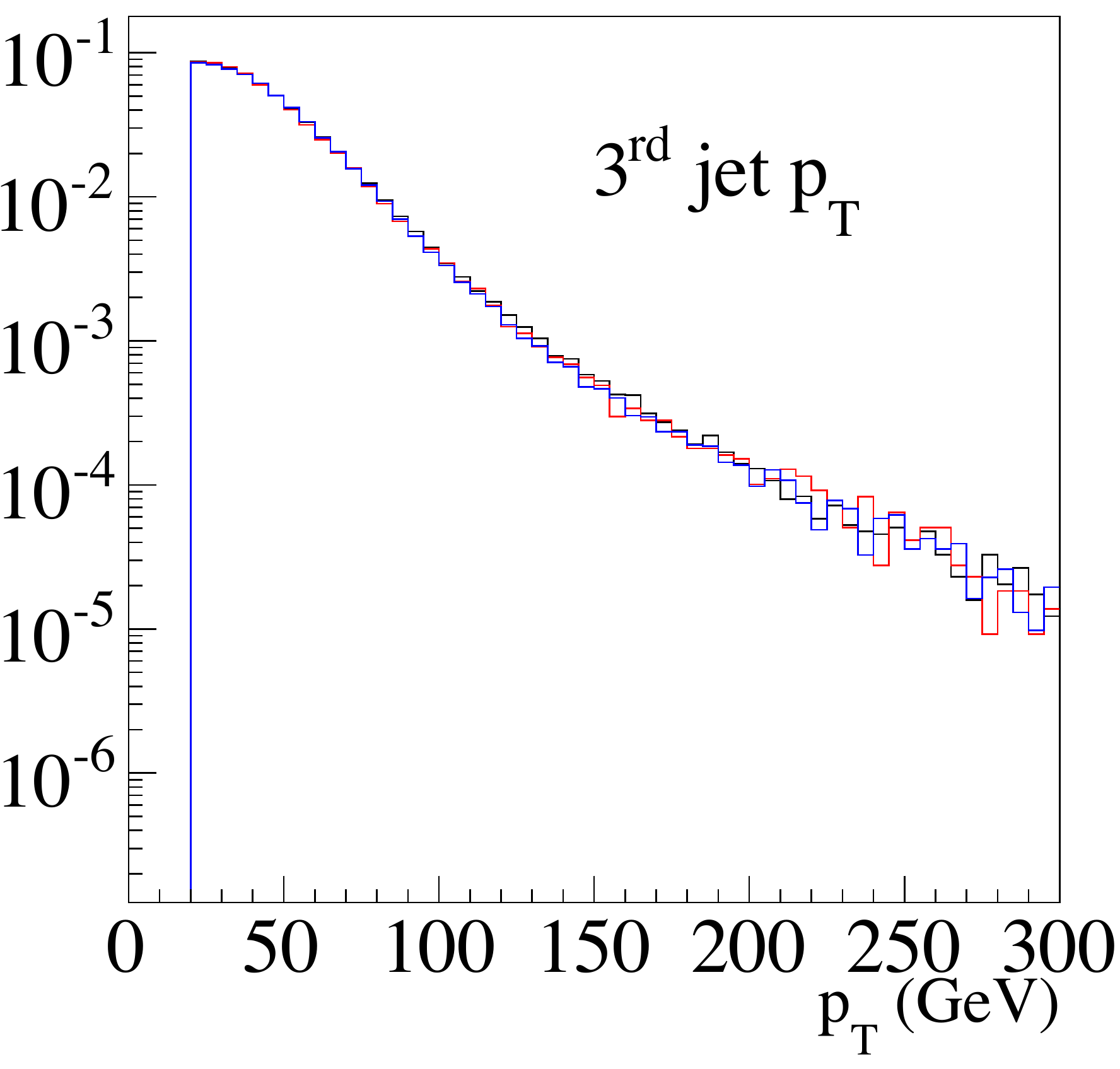}
\includegraphics[width=0.245\textwidth]{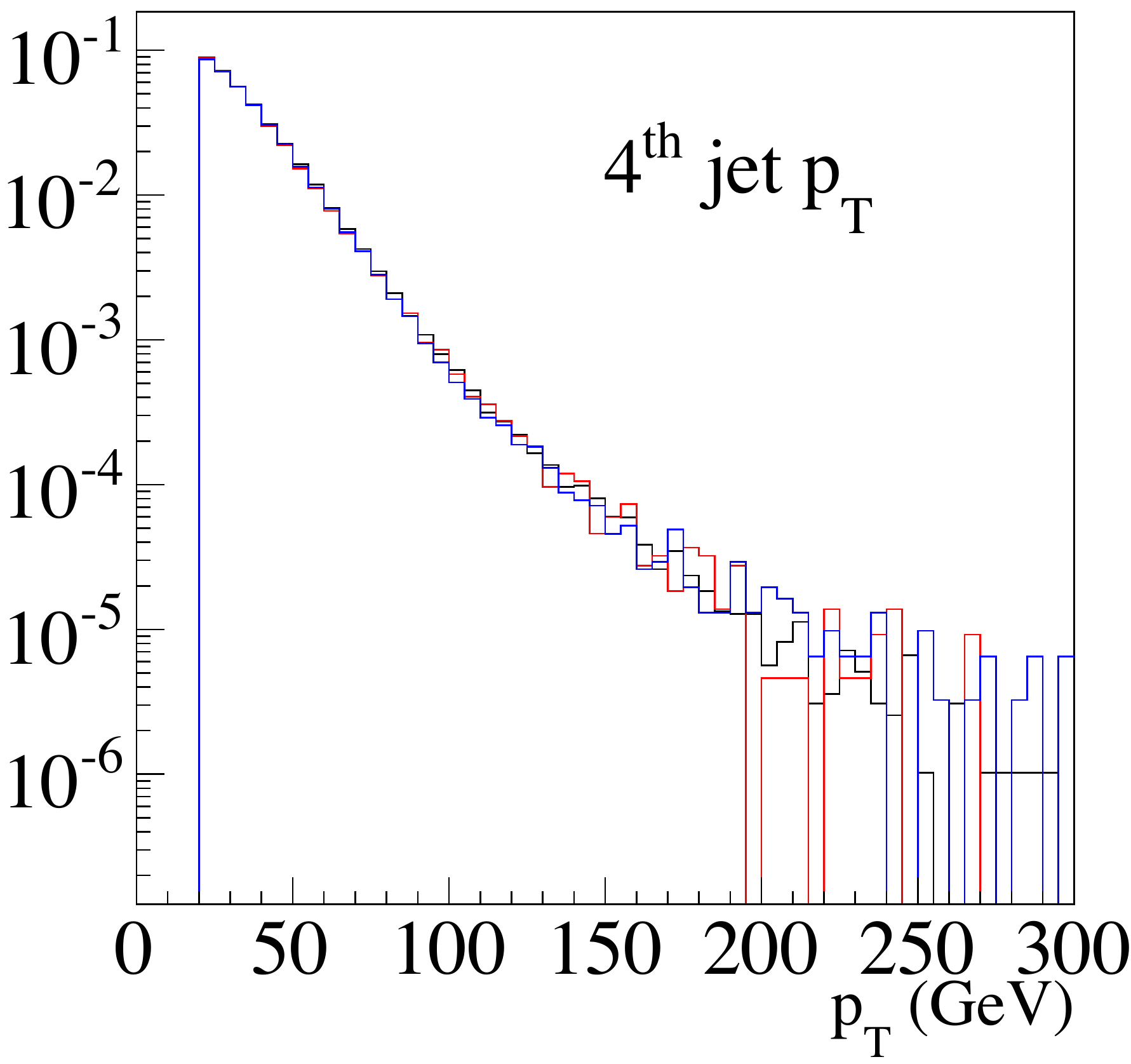}
\caption{Normalized $p_T$ distributions of the four leading jets in
  the merged $t\bar{t}$ samples, with {\tt xqcut}=20 (black), 40 (red),
  and 60~GeV (blue). We use anti-$k_T$ jets with $\Delta R = 0.5$,
  and require $p_T> 20$~GeV and $|\eta_j|<5$.}
\label{fig:qcut}
\end{figure} 
%---------------------------------------------

In order to extract information on the spin of the heavy top or stop
particles from tagging jets we need to ensure that our simulation
keeps all relevant spin correlations. Naively, this can be guaranteed by
generating events for the hard processes $\st{}\st{}^* jj$ and
$t\bar{t}jj$~\cite{skands,matt_michael,kaoru}.
However, the transverse momentum of the tagging jets will often be
significantly below the energy scale of this hard
process. In that region of phase space, for example the transverse momentum 
spectrum of jet radiation is only properly
described once we include the parton shower or other implementation of Sudakov factors. In standard
showering algorithms the
probabilistic parton shower is (usually) averaged over the helicities
of the participating partons. In such simulations, any apparent spin
correlation between the hard process and the tagging jets --or between
the tagging jets themselves-- comes only from kinematic
constraints~\cite{skands}, rather than from a combination of kinematics and underlying 
interference effects. What we need is a merged description
of the parton shower and the hard matrix element, where the tagging jets are
generated through the matrix element.\bigskip

To that end, we consider two benchmark parameter points for stop signals for stop pair
production followed by a decay into a top and a missing energy particle,
\begin{alignat}{5}
pp \to \st{} \st{}^* \to (t \nz{}) \; (\bar{t} \nz{}) 
\qquad \qquad 
(\mst,\mne{}) =
\begin{cases} (175, 1)~\gev \\ (200,25)~\gev \; .\end{cases}
\end{alignat}
The invisible particles coming from a
prompt decay can be a neutralino or a gravitino. As we are not
closely investigating the stop and top decay patterns we will refer to the
generic missing energy particle as $\nz{}$.

For the background and each signal benchmark we generate events for
the pair production of stops and tops at the 14 TeV LHC with up to
three extra jets in {\tt MadGraph5}~\cite{mg5,mlm}, matching the
jets to {\tt Pythia6}~\cite{pythia} and using anti-$k_T$
jets with $R=0.5$~\cite{fastjet} down to a matching scale {\tt
  xqcut}=20~GeV.  This choice (endorsed by the {\tt MadGraph} authors \cite{madgraphonline})
ensures that the spin correlations in
the tagging jets are kept, provided the two tagging jets are chosen
from the three leading jets that do not originate from top decay.  We will compare these results to
unmatched hard $t\bar{t}jj$ and $\st{} \st{}^*jj$
events~\cite{kaoru}. In this section we do not keep
track of the top and stop decays. The two tagging jets are the two
hardest jets which fulfill all $p_T$ and $\Delta \eta$
requirements.\bigskip

In order to ensure that all final state jets in {\tt MadGraph5} are
generated by the matrix element and hence include all spin and angular
correlations, we can move the matching scale to values below the
transverse momenta for all potential tagging jets, {\tt xqcut}$<
p_{T,j}$.\footnote{We have confirmed that for events with {\tt
    xqcut}$> p_{T,j}$ the correlations between the tagging jets in
  {\tt MadGraph} are indeed lost.}  While this choice will hugely
decrease the efficiency of the event generation, because a very large
fraction of events will be vetoed to generate the Sudakov suppression,
it will ensure that our events include all the necessary
information. Because the matching scale is not a physical parameter,
it can be varied within a reasonable range, where we will see that the
definition of `reasonable' is different for kinematic distributions
and the total rate.

%---------------------------------------------
\begin{figure}[t]
\includegraphics[width=0.245\textwidth]{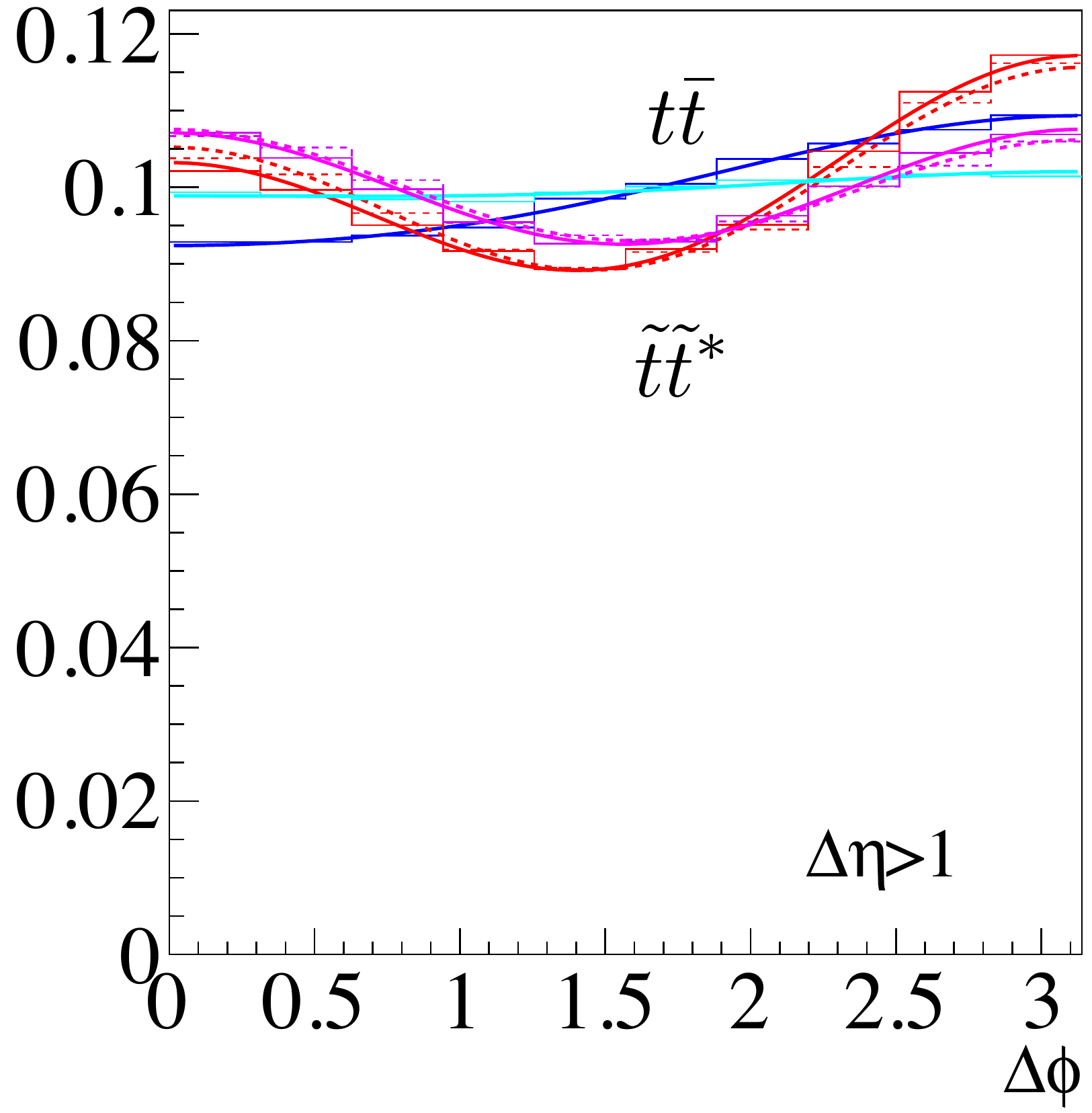}
\includegraphics[width=0.245\textwidth]{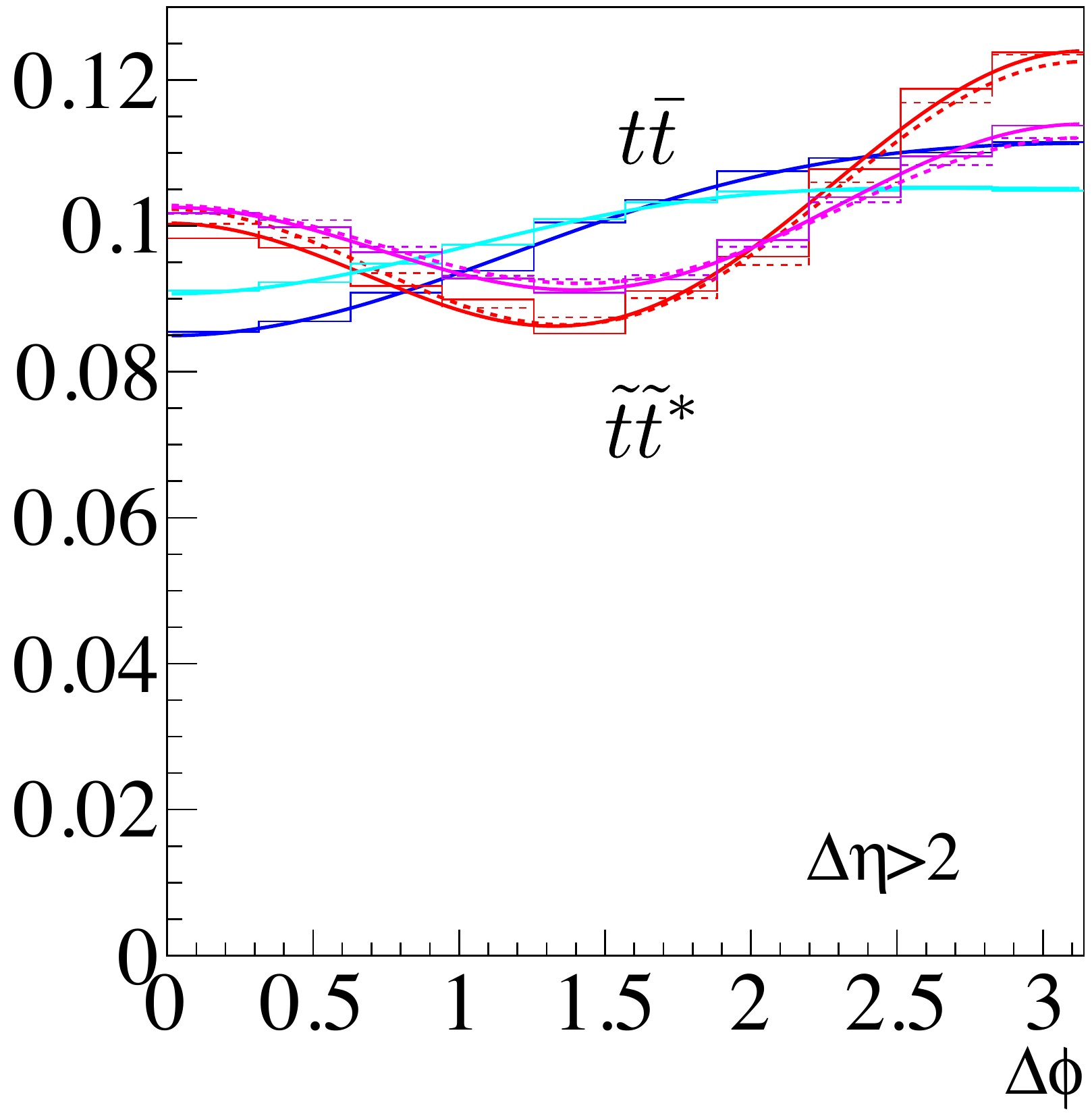}
\includegraphics[width=0.245\textwidth]{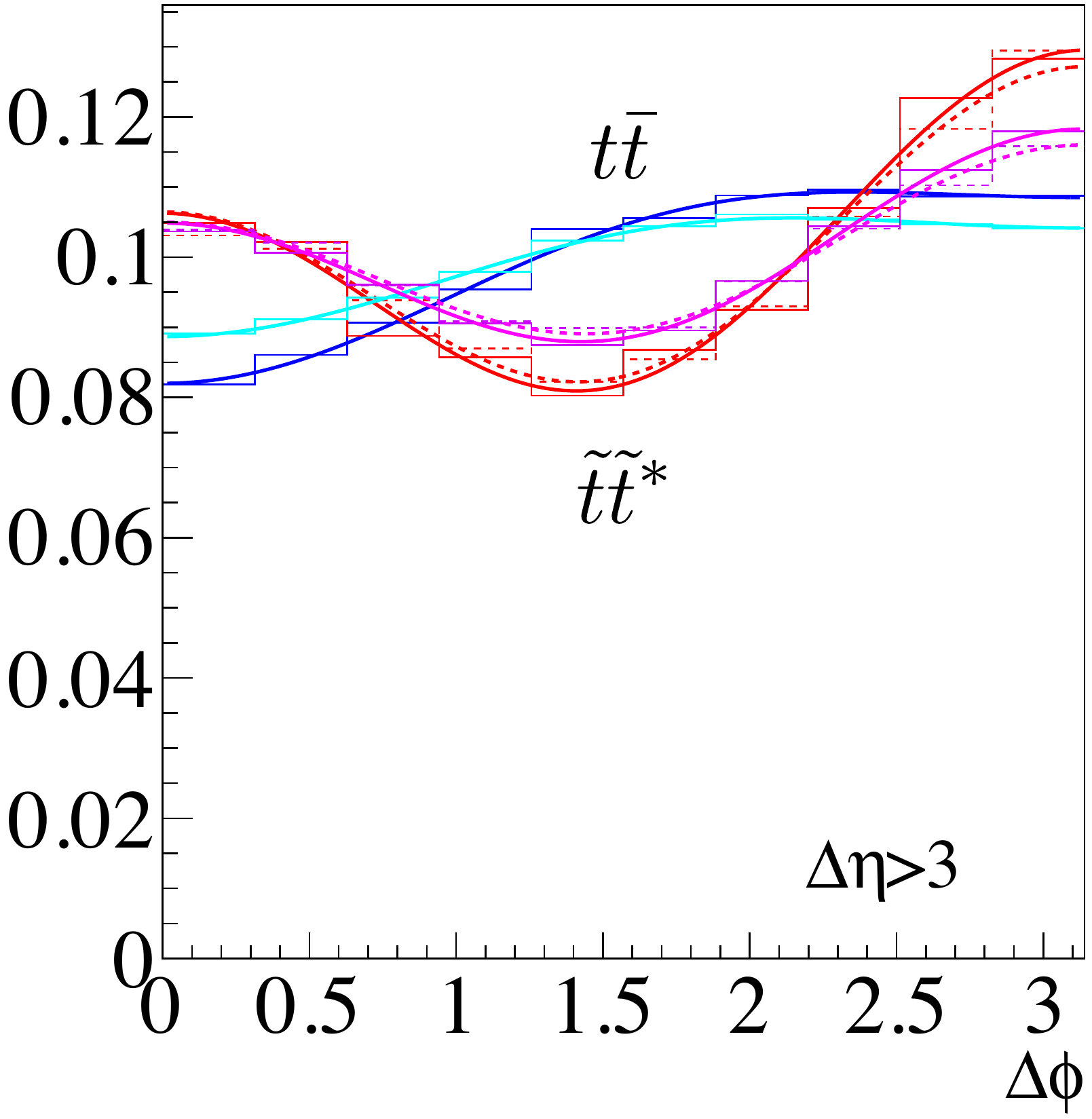}
\includegraphics[width=0.245\textwidth]{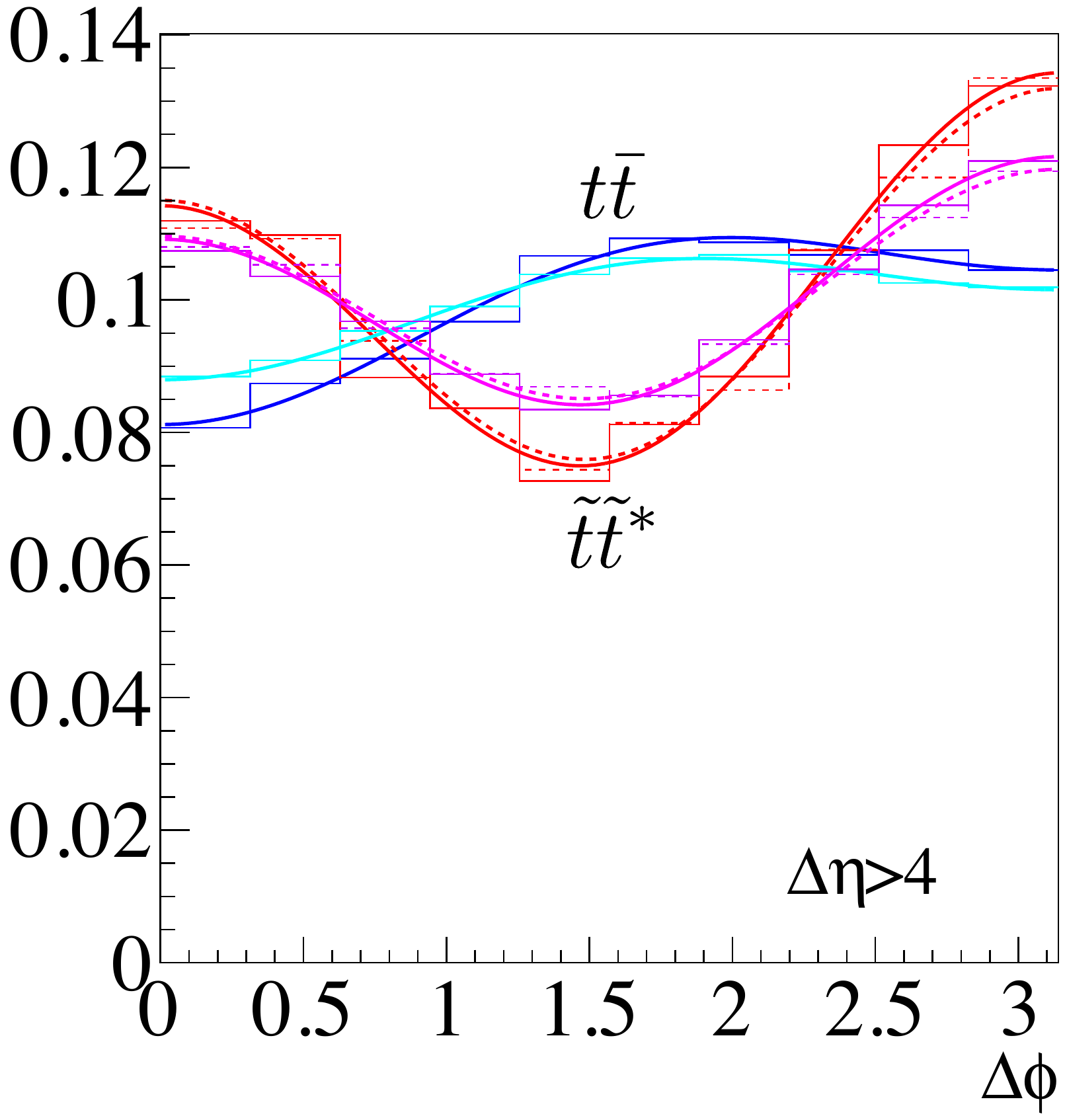}
\caption{Normalized $\Delta\phi$ distributions for the two
  highest-$p_T$ forward jets at parton level, requiring $\Delta \eta_{jj} >
  1,2,3,4$.  We show top pairs (blue) and stop pairs (red) matched to three jets,
  as well as the unmatched two-jet samples for tops (cyan) and stops (purple).  
   We also show the best fits to the functional form $A_0+A_1\cos\Delta\phi+A_2\cos
  (2\Delta\phi)$. For the stop samples, the $(\mst,\mne{}) = (175,1)$~GeV scenario is shown
  with a solid line, while $(\mst,\mne{}) = (200,25)$~GeV is shown
  with a dotted line.}
\label{fig:partondphi}
\end{figure}
%---------------------------------------------

Before we study the spin correlation between the tagging jets we test
if our choice of the matching scale, {\tt xqcut}=20~GeV,
leads to stable and consistent results.  To this end we show the
$p_T$ distributions for the first four jets for top pair production in
Figure~\ref{fig:qcut}. This distribution directly probes the Sudakov
suppression and should therefore be most sensitive to artifacts from
the choice of the matching scale. We vary the matching scale from
20~GeV to 40~GeV and the default value of 60~GeV. We see that the
distributions are essentially indistinguishable between the three
samples over the entire range of $p_T$, so our choice of scales does 
not present any problems for the tagging jet distributions.

On the other hand, the combined cross sections from {\tt MadGraph}
show a wider variation, with $\sigma_{t\bar{t}} = 2.9,~ 1.3,$ $0.94$,
and $0.71$~nb for {\tt xqcut}=20, 40, 60, and 100~GeV. Given that
multi-jet merging is based on a combination of leading order matrix
elements and a leading logarithmic parton shower, this variation
reflects the uncertainty of a leading order cross section with four
powers of $\alpha_s$. For smaller values of {\tt xqcut} we include
more and more real emission as described by the full matrix element,
but only compensated for by approximate virtual corrections in the
Sudakov factor. If we apply an external normalization of the total
production rate, for example to the precision predictions shown in
Table~\ref{tab:xsection} we can use a {\tt MadGraph} event samples
with the matching scale of 20~GeV to accurately simulate the
production of top or stop pairs plus jets.\bigskip

We can now consider the distribution of forward jets in top or stop events. 
In this Section, we will focus on confirming
the existence and the sign of the $A_2$ terms, as derived from the interference pattern described in 
Section~\ref{sec:spin}. Moreover, we need to test if our event generation 
indeed captures all relevant physics.
To be independent of the details of the top decay,
we use Monte Carlo truth to distinguish between associated
jets and those from top decay. For specific top decays it should be
straightforward to distinguish between ISR jets and decay jets, as has been shown
for direct production of supersymmetric particles~\cite{susy_isr}, for 
weak--boson--fusion pair production of supersymmetric
particles~\cite{wbf_isr}, and for sgluon pair
production~\cite{sgluon_isr}, as we will demonstrate shortly.  We then place selection criteria on
our 3-jet matched or 2-jet unmatched samples in order to isolate VBF-type production from all other
diagrams that generate two or more jets in association with stops or
tops.  Adapting the criteria used for WBF Higgs
selection~\cite{delta_phi,wbf_isr}, we begin by requiring at
least two parton--level jets in the merged sample with
\begin{equation}
p_{T,j} > 20~\gev, \qqqquad
|\eta_j|<5, \qqqquad 
\Delta \eta_{jj} > 1, 2, 3, 4 \; .
\label{eq:vbf_cuts}
\end{equation}
The increasing rapidity separation should emphasize the VBF-induced 
angular correlations between the tagging jets~\cite{higgs_spin}.
More realistic selection criteria will be put in place once we include
a fast detector simulation in Section~\ref{sec:searches}.\bigskip

%---------------------------------------------
\begin{table}[t]
\begin{footnotesize}
\begin{tabular}{ll|c|c|c|c|c|c|c|c} \hline
&  &  \multicolumn{2}{c|}{$|\Delta \eta_{jj}|>1$} & \multicolumn{2}{c|}{$|\Delta \eta_{jj}|>2$}  
   &  \multicolumn{2}{c|}{$|\Delta \eta_{jj}|>3$} & \multicolumn{2}{c}{$|\Delta \eta_{jj}|>4$}  \\ 
& & $A_1/A_0$ & $A_2/A_0$ & $A_1/A_0$ & $A_2/A_0$ & $A_1/A_0$ & $A_2/A_0$ & $A_1/A_0$ & $A_2/A_0$  \\ \hline 
\multirow{2}{*}{$t\bar{t}$} 
& 2-jet & $-0.016\pm0.03$ & $+0.005\pm 0.001$ & $-0.07\pm0.01$ & $-0.021\pm0.004$ & $-0.08\pm0.01$ & $-0.035\pm0.006$ & $-0.07\pm0.01$ & $-0.05\pm0.01$ \\
& 3-jet & $-0.08\pm0.01$ & $+0.009\pm0.002$  & $-0.13\pm0.02$ & $-0.018\pm0.003$ & $-0.13\pm0.02$ & $-0.048\pm0.008$ &  $-0.12\pm0.02$ & $-0.07\pm0.01$ \\ \hline
$\st{}\st{}^*$
& 2-jet & $-0.0023\pm0.0003$ & $+0.07\pm0.01$ & $-0.06\pm 0.01$ & $+0.08\pm 0.01$ & $-0.07\pm0.01$ & $+0.12\pm0.02$ & $-0.06\pm0.02$& $+0.15\pm0.02 $ \\ 
(175,1)
& 3-jet & $-0.07\pm0.01$ & $+0.10\pm0.02$ & $-0.12\pm0.02$ & $+0.12\pm0.02$ &  $-0.12\pm0.02$ & $+0.18\pm0.03$ & $-0.11\pm0.02$ & $+0.25\pm0.04$ \\ \hline
$\st{}\st{}^*$
& 2-jet & $+0.007\pm0.001$ & $+0.07\pm0.01$ & $-0.05\pm0.01$ & $+0.07\pm0.01$ & $-0.06\pm 0.01$ & $+0.11\pm 0.02$ & $-0.05\pm0.01$ & $+0.15\pm0.02$ \\ 
(200,25) 
& 3-jet & $-0.06\pm0.01$ & $+0.10\pm0.02$ & $-0.10\pm0.02$ & $+0.12\pm0.02$ & $-0.11\pm0.02$ & $+0.17\pm0.03$ & $-0.09\pm0.02$& $+0.24\pm0.04$ \\ \hline
\end{tabular}
\end{footnotesize}
\caption{Best-fit values for the $\cos\Delta\phi$ and $\cos
  (2\Delta\phi)$ coefficients
  defined in Eq.~\eqref{eq:diffsigma}. The fits are
  performed at parton level, corresponding to 
  Figure~\ref{fig:partondphi}. The 3-jet matched (2-jet unmatched) top background sample before any
  cuts consists of $1.95 \times 10^6$ ($3.09 \times 10^6$) events, the $\mst =175$~GeV
  stop sample is $5.65 \times 10^5$ ($6.18 \times 10^6$) events, and the $\mst =200$~GeV
  sample is $1.08\times 10^6$ ($9.24\times 10^5$) events.}
\label{tab:partonvbf}
\end{table}
%---------------------------------------------

In Figure~\ref{fig:partondphi} we plot the normalized $\Delta\phi$
distributions between the two highest-$p_T$ parton--level tagging jets
defined in the laboratory frame, requiring $\Delta \eta_{jj} > 1$,
2, 3, and 4 in the successive panels.  As can be seen, there is a clear difference between the
tagging jet correlations from stop and top events, corresponding to
the sign of the $\cos (2\Delta\phi)$ term.  It induces a clearly
visible minimum in the stop sample around $\Delta \phi = \pi/2$, especially
noticeable when compared to the slight excess here in the top sample.
Top pairs are dominated by a slight preference
for back-to-back tagging jets.

Without a $\Delta \eta_{jj}$ cut, the non-trivial azimuthal dependence
would be highly suppressed. This is expected, since central jets do
not predominantly come from the ISR diagrams and do not reflect
information about the helicity of fusing gluons through interference
patterns in our reference frame. As we enforce increasingly large $\Delta\eta_{jj}$ cuts we
see a finite $\cos (2\Delta\phi)$ component develop in both the top
and stop samples; with the appropriate signs for fermionic and scalar
pairs.
\bigskip

In Table~\ref{tab:partonvbf}, we show the relative size of the
$\cos\Delta\phi$ ($A_1$) and $\cos (2\Delta\phi)$ ($A_2$) modes for
the top background and stop benchmark points, normalized to the
constant term $A_0$. The coefficients are obtained from the normalized
ten--bin histograms at parton level, using the standard {\tt ROOT} fitting
algorithm. It is apparent that the non-trivial $A_2$ term is present
in the unmatched two-jet sample, and survives after the addition of a
third jet in the matching scheme. The magnitude of the $A_1$ term
significantly increases for the matched samples. 

Comparing the events with three merged jets and the events with
only two hard jets we see that the merged sample shows an
additional shift towards larger azimuthal tagging jet separation. The
reason is that with a third jet recoiling against the hard top or stop
pair system we now have a choice to pick the two tagging jets. We
systematically bias the selection towards an effectively larger
$\Delta \eta_{jj}$ separation translating into more back-to-back
tagging jets.  However, this shift mostly affects the $\cos \Delta
\phi$ distribution, while the critical $\cos (2\Delta\phi)$ mode is
symmetric around $\Delta \phi = \pi/2$ and therefore just slightly
tilted. The fact that for top pair production the kinematic effect
from additional jet radiation looks similar to the $\cos \Delta \phi$
mode from spin correlations explains the surprising finding of 
Ref.~\cite{skands} that the parton shower simulation seems to capture
some of the expected spin correlations while it should not.

The size of $A_2$ is only slightly affected by the
different simulational approaches shown in Table~\ref{tab:partonvbf}, \ie 
the theory-driven unmerged 2-jet setup and the more realistic merged
3-jet case. If anything, the effect in $\cos (2\Delta\phi)$ is more 
pronounced in the multi-jet case, contrary to what is observed as 
azimuthal decorrelation in 2-jet production. The two stop mass benchmarks are
consistent with each other.  Already for $\Delta \eta_{jj} >2$ we
observe the expected sign difference between the fermionic and scalar
processes. It will become an experimental issue how wide a
rapidity separation of the two tagging jets is needed to extract the
most information with a limited sample size.

%%%%%%%%%%%%%%%%%%%%%%%%%%%%%%%%%%%%%%%%%%%%%%%%%%%%%%%%%%%%%%%%%%%%%%%%%%%%%%%
\section{Stop Searches}
\label{sec:searches}  

%---------------------------------------------
\begin{table}[b!]
\begin{tabular}{ll|c|c|c|c}
\hline
 & & \multicolumn{2}{c|}{$|\eta_j|<2.5,~|\Delta \eta_{jj}|>2$} & \multicolumn{2}{c}{$|\eta_j|<4.5,~|\Delta \eta_{jj}|>3$} \\
 & & di-leptonic & semi-leptonic & di-leptonic & semi-leptonic \\ \hline 
\multirow{5}{*}{$t\bar{t}$} & leptons & 3.2\% & 29\% & 3.2\% & 29\%\\
 & +$b$-tag \& jets & 0.17\% &  0.98\% & 0.23\% & 1.5\%\\
  & +$W$-mass & -- & 0.19\% & --  & 0.25\%\\
 & +$|\Delta \eta|$ & 0.053\% & 0.066\% & 0.061\% & 0.064\%\\
 & Final $\sigma$ & 505~fb & 629~fb & 582~fb & 610~fb\% \\ \hline
\multirow{5}{*}{$\st{} \st{}^*$ (175,1)} & leptons & 3.3\% & 29\% & 3.3\% & 29\%\\
 & +$b$-tag \& jets & 0.14\% & 0.87\% & 0.19\% & 1.3\%\\
 & +$W$-mass & -- & 0.17 \% & -- & 0.23\%\\
 & +$|\Delta \eta|$& 0.041\% & 0.060\% & 0.048\% & 0.058\% \\
 & Final $\sigma$  & 55~fb & 81~fb & 65~fb & 78~fb \\ \hline
\multirow{5}{*}{$\st{} \st{}^*$ (200,25)} & leptons & 3.3\% & 29\% & 3.3\% & 29\%\\
& +$b$-tag \& jets & 0.17\% & 1.1\% & 0.23\% & 1.6\%\\
 & +$W$-mass & -- & 0.22\% & -- & 0.28\%\\
 & +$|\Delta \eta|$ & 0.050\% & 0.076\% & 0.057\% & 0.069\%  \\ 
 & Final $\sigma$ & 36~fb & 55~fb & 41~fb & 50~fb \\ \hline
\end{tabular}
\caption{Cumulative efficiencies, including branching
  ratios, after detection selection criteria, in both di- and
  semi-leptonic channels. Also shown is the cross section
  after all cuts are applied. The ``leptons'' cut requires two (one) $e$ or $\mu$
  for the di-lepton (semi-leptonic) channel. Two $b$-tagged and two (four)
  or more non-$b$-tagged jets are required to pass ``$b$-tag \& jets,'' and the semi-leptonic
  $W$-mass reconstruction is defined in the text. The final $|\Delta \eta|$ criteria
  is applied for both jet selection criteria as defined
  in Eq.~\eqref{eq:jet_def}.}
\label{tab:efficiencies}
\end{table}
%---------------------------------------------

The results obtained in the last section at parton level and using Monte--Carlo truth
clearly demonstrate the analytic argument of Section~\ref{sec:spin}.
Once all helicity information is taken into account and kinematic cuts
restrict events to the VBF phase space, the stop events have a
positive coefficient $A_2$, while the top background has a negative
$A_2$. However, these results do not yet demonstrate that this
difference between scalars and fermions can be used to enhance the
stop sample among tops in a real experiment. One might worry that the identification of the 
tagging jets, combinatorics, or detector effects could wash out
these correlations and make them experimentally invisible.\bigskip

To confirm the experimental accessibility of the azimuthal correlation
as a way to separate top pairs from stop pairs we now hadronize the
parton level event samples with {\tt Pythia} and apply the fast
detector simulation {\tt Delphes3}~\cite{delphes} with
configuration files provided by the Snowmass Energy Frontier
simulations~\cite{snowmass}. Jets
are clustered using the anti-$k_T$~\cite{fastjet} algorithm
with $R = 0.5$.  All decays are included via {\tt Pythia}, so we do
not systematically account for spin correlations and interference
patterns in the production and decay processes. From the last section
it is clear that the details of the top and stop decays play no role
in our analysis, beyond triggering and combinatorial challenges. In
our analysis we include both semi-leptonic and di-leptonic top pair
decays. Fully hadronic decays of tops could be added once we resolve
QCD and combinatorical issues, discussed for example in
Refs.~\cite{combinatorics}.

We generate the equivalent of 4.8~fb$^{-1}$ of 14 TeV LHC data for the
top background and both stop signal points. Although this is much less than the
planned integrated luminosity of the next stage of LHC running, generating the corresponding full data set
would be extremely resource intensive and not essential for purposes of demonstrating
the feasibility of the $\Delta\phi$ technique. Indeed, as we will show below, even
with only $\sim 5$ fb$^{-1}$, the interference effect can already make stops known
in the top sample, though additional luminosity would be required to improve
the statistical significance.\bigskip

Depending on the assumed decay channel we require one or two electrons
and muons, required to have 
\begin{alignat}{5}
p_{T,\ell} > 20~\gev \quad \text{and} \quad |\eta_\ell|<2.5 \; . 
\end{alignat}
Regardless of the selection criteria of forward
jets, we require exactly two  $b$-tagged jets 
with 
\begin{alignat}{5}
p_{T,b}> 50~\gev \quad \text{and} \quad  |\eta_b|<2.5 \; ,
\end{alignat}
using the {\tt Delphes3} efficiency of approximately 70\% per $b$-tag.
For the upcoming 14~TeV runs of the LHC, where pile-up and jet energy
calibration might be an issue, we follow two potential choices for the
jet requirements,
\begin{alignat}{5}
(1) \qquad p_{T,j} &> 20~\gev \quad \mbox{and} \quad |\eta_j| <2.5 \notag \\
(2) \qquad p_{T,j} &> 20~\gev \quad \mbox{and} \quad |\eta_j| <4.5 \; .
\label{eq:jet_def}
\end{alignat} 
While the conservative assumption will prove to be sufficient to
reveal the presence of degenerate stops, including tagging jets to
$|\eta|<4.5$ will improve the physics reach in
this type of search.

For the di-leptonic channel, we require two or more light-flavor
jets. In the semi-leptonic channel we require four or more jets. Due
to limited statistics, in the di-leptonic channel we do not subdivide the events 
into different lepton flavor
combinations, though this could be useful for a full experimental
analysis.  Similarly, a full experimental analysis might find it
useful to include a systematic multi-jet analysis for tagging jets as well
as decay jets~\cite{moments_wbf},
but in this paper we limit ourselves to the cleanest possible
signature.

To differentiate the $W$-decay jets from the VBF tagging jets in the
semi-leptonic channel, we suggest the following reconstruction
algorithm: of all pairs of central ($|\eta_j|<1$) jets passing
a staggered cut $p_{T,j} > 60,30$~GeV we take the pair with an invariant mass
closest to $m_W$.  If an event has such a pair of jets and their
invariant mass is within 30~GeV of the $m_W$, it is retained for the
VBF selection criteria. The two highest-$p_T$ QCD jets remaining must then have an
invariant mass of either less than 50~GeV or greater than 100~GeV, to
avoid possible misidentification with the $W$-boson decay
products. This strict set of requirements provides a very clean sample
of events where the two VBF jets are well separated from all other
hadronic activity in the detector, though the efficiency is
correspondingly low, and improvements on this algorithm are obviously possible.

The highest-$p_T$ non-$W$-tagged jets in the semi-leptonic sample and the
highest-$p_T$ jets in di-leptonic events are likely be the two tagging jets, so we apply
the $\Delta\eta_{jj}$ cut. In the conservative jet selection
scenario~(1) with $|\eta_j|<2.5$ we only require $|\Delta \eta_{jj}| >
2$, in order not to cut too deeply into the efficiency. For the more
optimistic situation~(2) with $|\eta_j|<4.5$ we can also require a
larger jet separation: $|\Delta \eta_{jj}| > 3$.  From all events
passing this final cut we construct the $\Delta\phi$ distribution. The
final efficiencies and effective cross sections for both the di- and
semi-leptonic channels are shown in
Table~\ref{tab:efficiencies}, including the efficiencies of each cut
leading up to the final $\Delta \eta$ selection.\bigskip

%---------------------------------------------
\begin{table}[t]

\begin{tabular}{cc|c|c|c|c} \hline
 & &  \multicolumn{2}{c|}{$|\eta_j|<2.5,~|\Delta \eta_{jj}|>2$} & \multicolumn{2}{c}{$|\eta_j|<4.5,~|\Delta \eta_{jj}|>3$}  \\ 
 & & di-leptonic $A_2/A_0$ & semi-leptonic $A_2/A_0$ & di-leptonic $A_2/A_0$ & semi-leptonic $A_2/A_0$  \\ \hline 
\multicolumn{2}{c|}{$t\bar{t}$} & $-0.10 \pm 0.03$ & $-0.05\pm 0.03$ & $-0.12\pm 0.03$ & $-0.08 \pm 0.03$ \\ \hline
\multirow{2}{*}{$\st{} \st{}^*$ (175,1)} & $\st{} \st{}^*$ only & $+0.20\pm0.09$ & $+0.10\pm0.07$ & $+0.16\pm0.09$ & $+0.18 \pm 0.07$ \\ 
 & $\st{} \st{}^* + t\bar{t}$ & $-0.07\pm0.03$ & $-0.03\pm0.02$& $-0.09 \pm 0.03$& $-0.05\pm 0.02$ \\ \hline
\multirow{2}{*}{$\st{} \st{}^*$ (200,25)} & $\st{} \st{}^*$ only &  $+0.22\pm0.11$ & $+0.03\pm0.08$ & $+0.18 \pm 0.11$ & $+0.16\pm 0.10$ \\
 & $\st{} \st{}^* + t\bar{t}$ & $-0.08\pm0.03$ & $-0.04\pm0.01$ & $-0.10 \pm 0.03$ & $-0.06\pm 0.03$ \\ \hline
\end{tabular}

\caption{Best-fit values for the $\cos
  (2\Delta\phi)$ coefficients $A_2$, normalized to the constant term $A_0$,
  defined in Eq.~\eqref{eq:diffsigma}, for di-leptonic and semi-leptonic events
  corresponding to 4.8~fb$^{-1}$ of luminosity, after fast detector simulation. 
%  The fits are shown for both the conservative and optimistic jet selections.
  Fits to the two stop signal points are performed for signal only as well as
  signal plus top background.}
\label{tab:delphesvbf}
\end{table}
%---------------------------------------------

Based on the 4.8~fb$^{-1}$ of simulated signal and background data
 given in Table~\ref{tab:delphesvbf}, we can extrapolate
what integrated luminosities would be required to observe a significant number
of stop pair events inside the top sample. Clearly, the statistical errors from 5~fb$^{-1}$
of integrated luminosity would be too large to make any statement,
as the difference between the background distribution and the background
plus signal is equivalent to the fit uncertainties. 

However, by taking the central fit values of the $d\sigma/\Delta\phi$ differential
distribution as the `true' parameter values, we can determine the statistical 
power for a given amount of data. 
The luminosity from the first year of LHC14 running is expected to be around 25~fb$^{-1}$. 
This data set would reduce the statistical errors on the $A_2/A_0$ parameter to 
approximately $1\%$. This would allow a $\sim 1.5\sigma$ statistical
differentiation between background and background plus signal for 175~GeV
tops in the di-leptonic channel ($\sim 1\sigma$ for 200~GeV stops) in the current detector 
configuration, and somewhat less in the semi-leptonic channel. With improved jet tracking in the forward region, this might be
improved to $1.7\sigma$ with a year's luminosity. With a data set of 100~fb$^{-1}$, 
$3.2\sigma$ observation would be possible in both channels for 175~GeV
stops, and $2\sigma$ discovery for 200~GeV stops, assuming the
conservative $|\eta|$ requirements. This would be improved to
$3.7\sigma$ for 175 GeV ($2.4\sigma$ for 200 GeV) stops assuming the
detector performance allows for $|\eta| <4.5$ in the tagging jets.

Such statements do not include systematic errors, which are clearly of concern for an observable 
so dependent on jet reconstruction and identification. However, analysis of tagging jets
has already been proven to work in Higgs studies with the 8~TeV run.
Moreover, as noted in this paper, several handles are available to allow experimental 
control of these issues. The signal will be visible in 
both semi-leptonic and di-leptonic decays, and with sufficient luminosity the
di-leptonic channel could be further broken down into the different flavor
combinations. The turn-on of the non-trivial $A_2$ signal as the $\Delta \eta$
cut is instituted provides an important cross check, and it is possible
that selection cuts intended to isolate the $\beta$ dependence~\cite{kaoru} of the
top and stop signals will also define useful side-bands.

%%%%%%%%%%%%%%%%%%%%%%%%%%%%%%%%%%%%%%%%%%%%%%%%%%%%%%%%%%%%%%%%%%%%%%%%%%%%%%%
\section{Conclusions}
\label{sec:conclusion}

During the first LHC run tagging jets have been shown to be powerful tools
in observing Higgs decays to photons, $W$-boson, and tau-leptons. In the 
coming LHC runs with almost twice the collider energy their role will become
even more pronounced, also reaching beyond Higgs analyses.
Similar to the spin and CP studies based on weak--boson--fusion Higgs
events~\cite{delta_phi,higgs_spin}, we can test top quark properties
in top pair production with two forward
jets~\cite{kaoru,matt_michael}. This tagging jet analysis has the general 
advantage that it does not rely on the reconstruction of the hard process, in our
case the top pair. Instead, we can use the dependence on the azimuthal angle
$\Delta\phi$ between the tagging jets to search for non-standard events in the top
sample at the LHC. Specficially, the coefficient $A_2$  of the $\cos (2 \Delta \phi)$ 
term in the distribution is negative for top pair production, whereas 
light scalar top pairs will give a significant positive
contribution to this observable.\bigskip

We first showed how the different signs can be understood in terms of
the gluon helicity combinations contributing to the total rate. We
then established and tested a non-standard {\tt MadGraph5} setup which
allows us to simulate events with all angular correlations between the
ISR tagging jets intact. Using this modified generation tool we showed
that the precision on the extraction of $A_2$ increases with the
rapidity separation of the tagging jets. We also saw that the $A_2$
mode is not sensitive to the details of the ISR tagging jet simulation
and the model parameters in the stop decays. Finally, we estimated
that such an analysis should give $>3\sigma$ results in multiple channels
with around 100~inverse femtobarns of data at a 14~TeV LHC. Because the analysis is purely
based on the tagging jets is can be generalized to any hard process in
and beyond the Standard Model.\bigskip

\begin{center}
{\bf Acknowledgments}
\end{center}

We would like to thank Stefan Prestel for checking that our {\tt
  MadGraph5} simulation makes sense. MB would like to
thank Maria Spiropulu, Joe Lykken, Yuri Gershtein, and John-Paul Chou
for helpful discussion and resources.  MB and MJRM thank the Aspen
Center for Physics, where this project was originally conceived,
while TP foolishly skipped the workshop. Finally, TP would like to 
thank Frank Krauss for deep insights into azimuthal decorrelation. This work was supported in part by U.S. Department of Energy contract DE-SC0011095 (MJRM).

%%%%%%%%%%%%%%%%%%%%%%%%%%%%%%%%%%%%%%%%%%%%%%%%%%%%%%%%%%%%%%%%%%%%%%%%%%%%%%%

\end{document}